\documentclass[twocolumn,preprintnumbers,amsmath,amssymb,prb]{revtex4-2}
\usepackage{graphicx} 
\usepackage{amsmath}
\usepackage{amssymb}
\usepackage{multirow}
\usepackage{appendix}

\newcommand{\ket}[1]{|#1\rangle}
\newcommand{\bra}[1]{\langle#1|}

\begin{document}
\title{Entanglement in a molecular Lieb-lattice quantum computing circuit: A tensor network study}
\author{Wei Wu}\email{wei.wu@ucl.ac.uk}

\affiliation{UCL Department of Physics and Astronomy and London Centre for Nanotechnology, University College London, Gower Street, London WC1E 6BT}

\begin{abstract}
Here a finite-Lieb-lattice quantum computing circuit consisting of spin-$\frac12$ quantum bits (qubits) and triplet couplers is designed. Important gradient - quantum entanglement - is analyzed. This type of design could be realized in a vast range of molecules containing multiple radicals, in which the communications among qubits are controlled by the optically driven triplets. The von Neumann entanglement entropy, reduced density matrices, and spin-spin correlations were computed using tensor-network methods by varying the magnetic anisotropy and external magnetic field. At low magnetic fields and anisotropies, the bipartite entanglement entropies reach the maximum at the edge. By contrast, the entanglement peak will shift toward the bulk as the external magnetic field and the magnetic anisotropy increase, owing to a quantum phase transition. The reduced density matrices and spin–spin correlations reveal a mesoscopic entanglement across the molecular network. This work uncovers the rich entanglement patterns, quantum phase transitions, and tunable spin coherence in this mixed spin system, designed for molecular spin-based quantum computing. These findings have important implications for triplet-mediated molecular self-assembly quantum computing circuit, especially for the entangling gate based on molecules. This work would provide a theoretical cornerstone for the experimental realization of scalable molecule-based quantum computing circuits.
\end{abstract}

\maketitle

\section{Introduction}



As quantum science and technology shape the future, the development of efficient quantum computing circuit architectures will be of central importance \cite{alexeev2021quantum}. Molecular-spin-based quantum circuits hold great potential for scalable QC by leveraging the intrinsic quantum nature of molecules and their self-assembly functionalities for a molecular network \cite{wan2006fabricating,otero2011molecular,zhuang2015two}. Molecular networks could be chemically tailored to host both spin quantum bits (qubits) and couplers to form QC circuit \cite{wu2025stable}. With precise control over spin coherence and coupling, molecular quantum circuits could enable scalable and programmable quantum information processing in a customizable architecture \cite{gaita2019molecular,wu2025stable}. The vast-scale molecular database offers unparalleled versatility in the design of QC circuits. Numerous recent developments in the spin-based molecular qubits such as the ultra-long life time in covalent organic frameworks (COF) recently found illustrate the potential of building qubits and couplers  into the molecular network - a molecular QC circuit \cite{warner2013potential,sun2025ultralong, yamauchi2024toward}.

Control of the coupling between molecular spins is of paramount importance for the realization of quantum gate operations. One of the experimentally achievable routes to mediate the interaction between molecular spins could be the optically driven molecular triplet, which is well understood and can be used to reduce decoherence \cite{quintes2023properties,huang2023triplet,wu2025stable}. In this coupling mechanism, the optically addressable triplet coupler can mediate the spin-spin interaction between the spin-$\frac{1}{2}$ qubits, stemming from the triplet-radical system (TRS) \cite{quintes2023properties}. The ongoing advances in the chemical synthesis and electron spin resonance are pushing TRS closer to practical quantum gate operations \cite{gorgon2023reversible}. 

\begin{figure}[htbp]
\centering
\includegraphics[scale=0.225, trim={0.cm 0cm 0.0cm 0.0cm},clip]{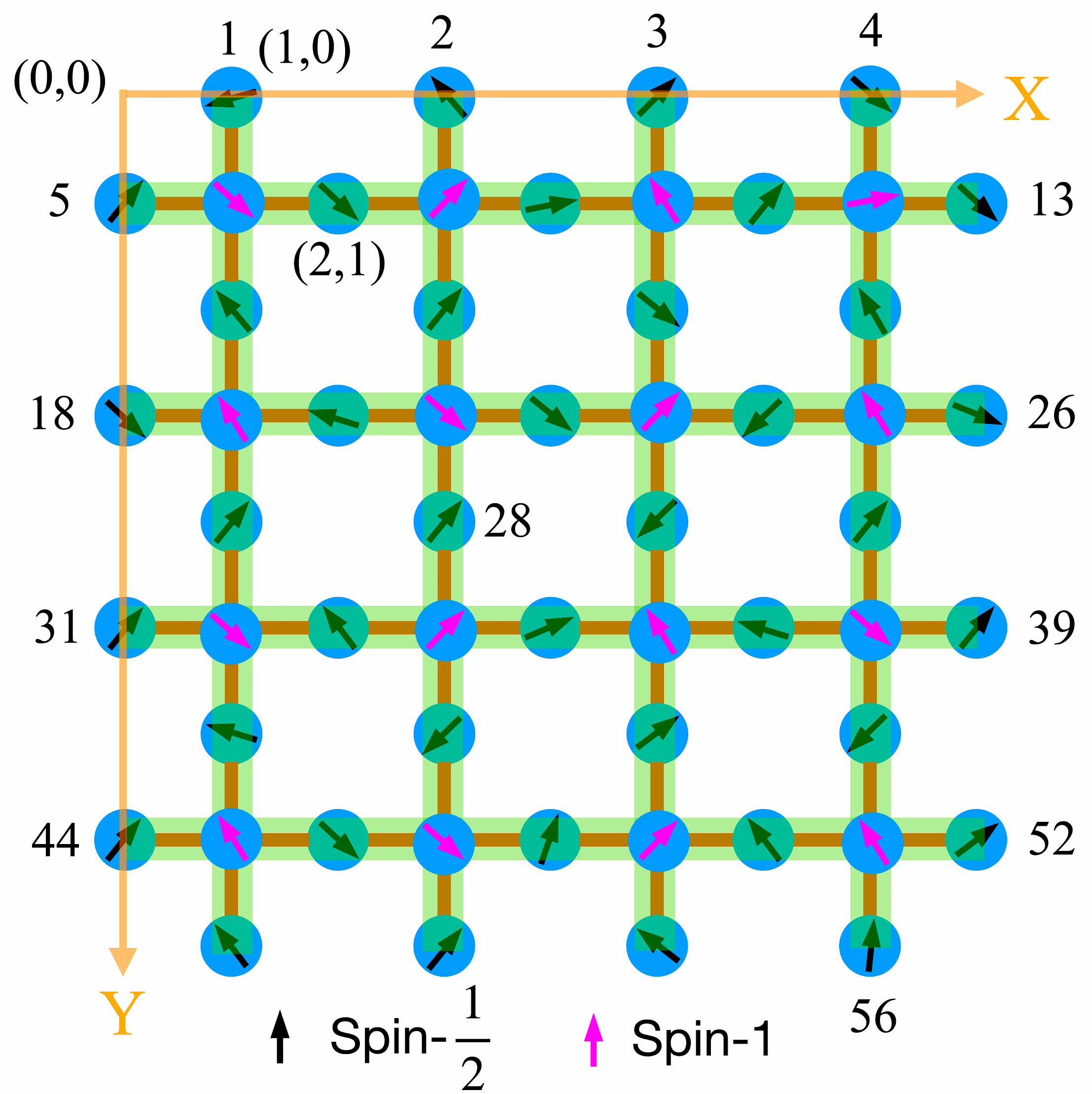}\\
\caption{The molecular QC network consisting of 40 $\frac12$-spins and 16 triplets is shown. The first spin ($s=\frac12$) has a coordinate of (1,0). And the spins are ordered by rows from left to right, starting from the top left corner. Each optically driven triplet is surrounded by four qubits, mediating their communications.}\label{fig:molecularqc} 
\end{figure}

Here, based on the previous pioneering QC circuit based on the superconducting qubits, a molecular QC network (Fig.\ref{fig:molecularqc}) is designed by using spin-$\frac12$ (qubit) and triplet spin coupler \cite{Arute2019,wu2025stable}. The spin-$\frac12$ could be carried by spin-bearing organic molecules such as radicals, while the triplet coupler could be closed-shell molecules with a good optical response \cite{wu2025stable}. Each triplet coupler is surrounded by four qubits. This two-dimensional (2D) molecular network can also be seen as a finite-size Lieb lattice \cite{slot2017experimental}.   
For this type of molecular QC network, the knowledge gap lies in the understanding of the quantum entanglement (the essential ingredient for QC) in this molecular quantum circuit and its feasibility for QC.  Moreover, the underlying physics for the quantum state and entanglement in this type of mixed-spin network is also of great interest for quantum spin models. In light of these, it would be crucial and timely to theoretically analyze the entanglement structure for the molecular circuit design.  

In this report, the von Neumann entanglement entropy, the key reduced density matrix elements for the spin coherence, and the spin-spin correlation functions have been analyzed using the tensor network techniques. The effects of the magnetic anisotropy of triplet and the external magnetic field on the entanglement entropy and the reduced density matrix elements have been studied. The edge mode of the entanglement entropy and the long-range entangled spins mediated by the triplets have been illustrated. The results here provide a new insight to the molecular spin-based quantum entanglement in a mixed spin system.




\begin{figure}[htbp]
\centering
\includegraphics[scale=0.135, trim={1cm 3cm 0.0cm 1.5cm},clip]{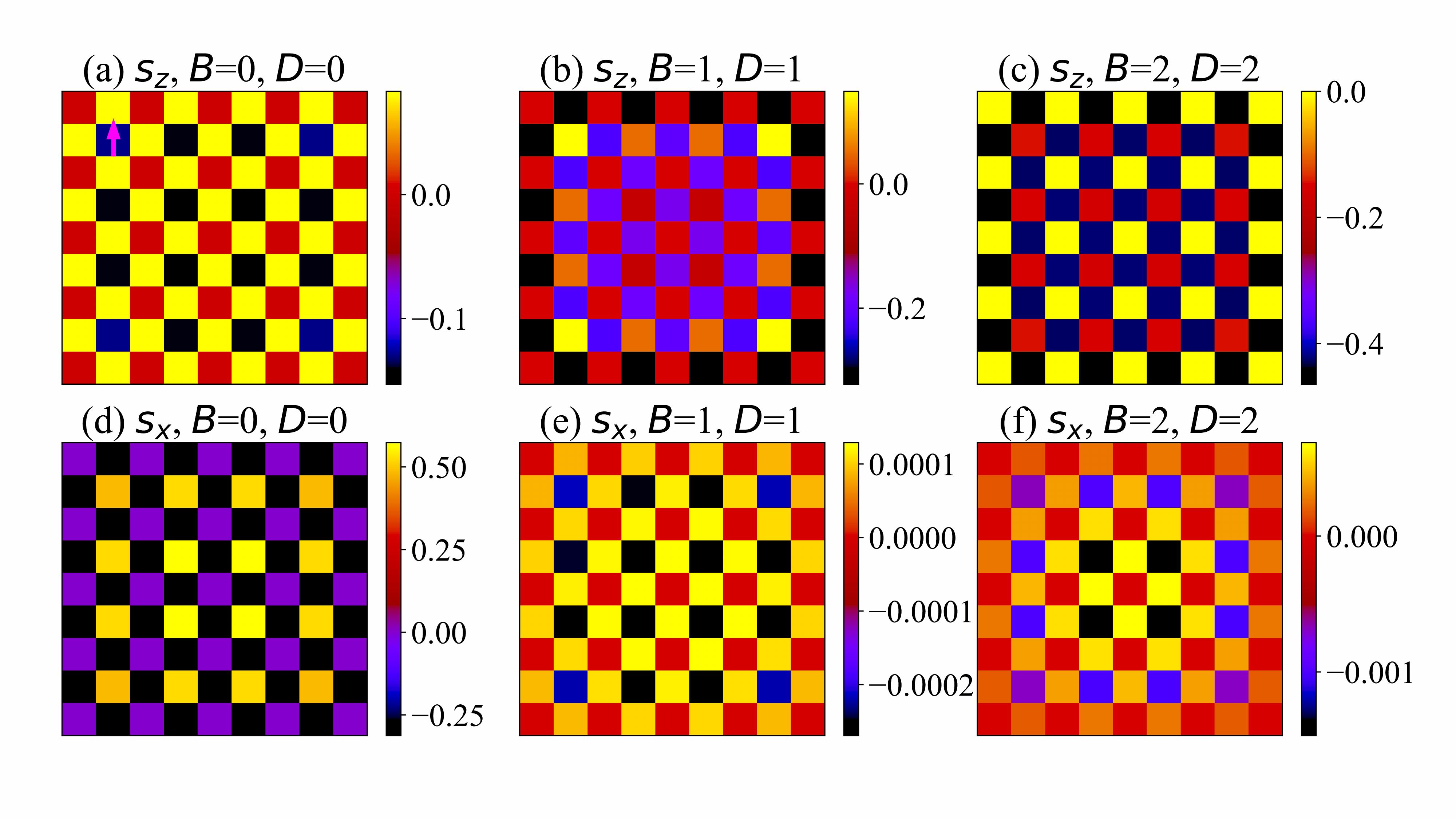}\\
\caption{The expectation values of $\hat{s}_{x,z}$ and $\hat{T}_{x,z}$ for the ground state are shown. When $B$ and $D$ are small, the triplets are anti-aligned with the surround $\frac12$-spins. When $B$ and $D$ get larger, the spins start to get aligned on the edge. The first triplet is labeled with a pink arrow to guide the eye.}\label{fig:sxsz} 
\end{figure}

\begin{figure}[htbp]
\centering
\includegraphics[scale=0.16, trim={5.5cm 3cm 0.0cm 2.0cm},clip]{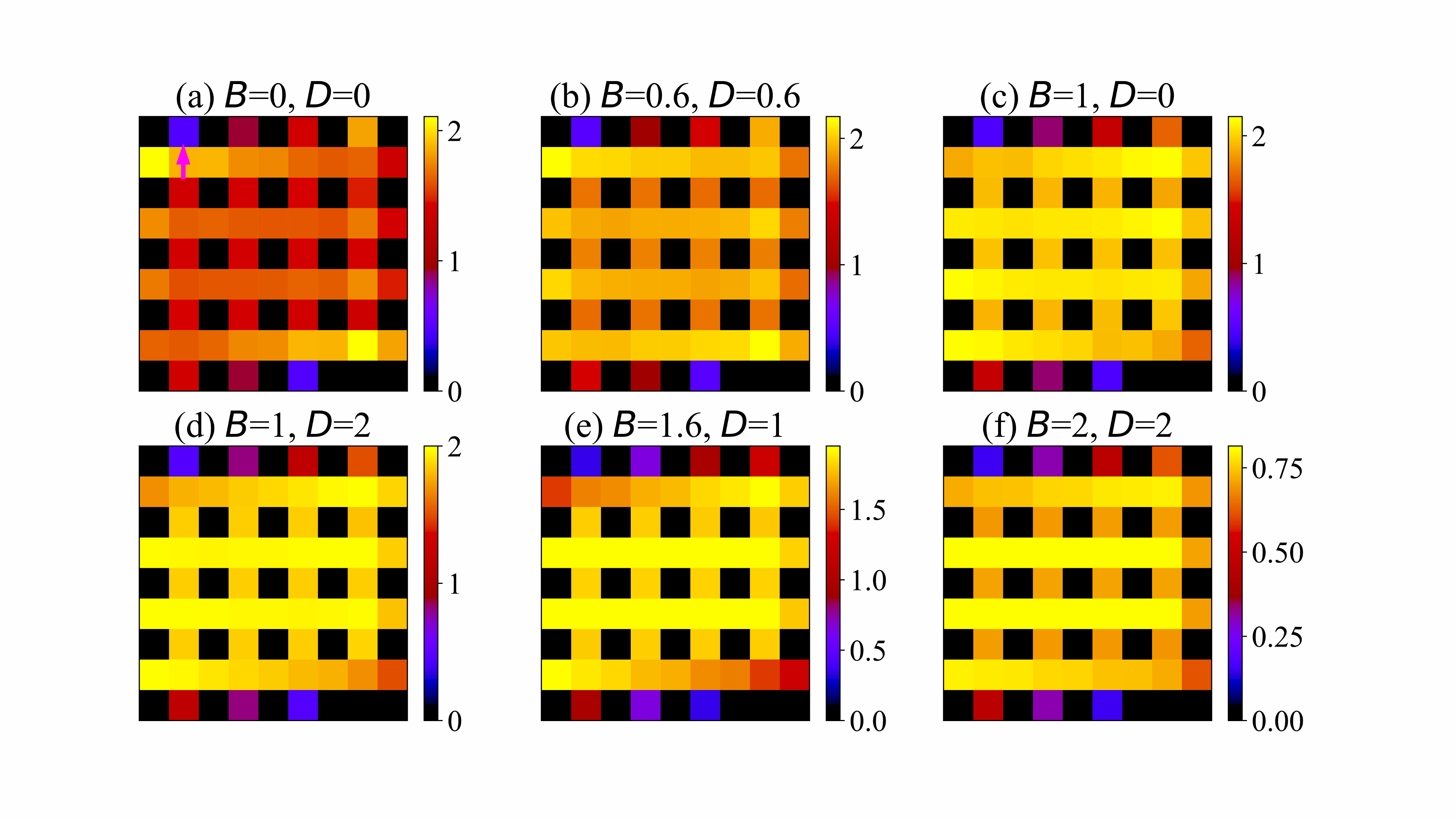}\\
\caption{The entanglement entropy map for the molecular network shown in Fig.\ref{fig:molecularqc}. The color-coded square at the site $n$ represents the bipartite entanglement entropies between the first $n$ spins and the remaining ones. The first triplet is labeled with a pink arrow to guide the eye.}\label{fig:ent1} 
\end{figure}

\section{Results and Discussions}
The bipartite entanglement values for the quantum circuit are shown in Fig.~\ref{fig:ent1} (Please find more details for the spin Hamiltonian defined in eq.\ref{eq:1}). The site $n$ is color-coded according to the value of the entanglement entropy between the first $n$ spins and the remainder of the system. When $B$ and $D$ are weak compared with the exchange interaction ($B=D=0$ or $0.6$), the entanglement entropy peaks at the fifth spin (a spin-$\frac{1}{2}$), indicating that the spins along the upper edge and the fifth spin exhibit the strongest bipartite entanglement with the remaining. This edge-dominant entanglement is suppressed when the relatively strong external magnetic field and single-ion anisotropy are introduced. Under these conditions, the entanglement entropy becomes more pronounced in the bulk of the circuit. Thus, the magnetic field and anisotropy act to both enhance and shift entanglement toward the center of the network. Therefore there could be a quantum phase transition in between when varying $B$ and $D$.

\begin{figure}[htbp]
\centering
\includegraphics[scale=0.24,trim={0.5cm 0.5cm 0.0cm 0.5cm},clip]{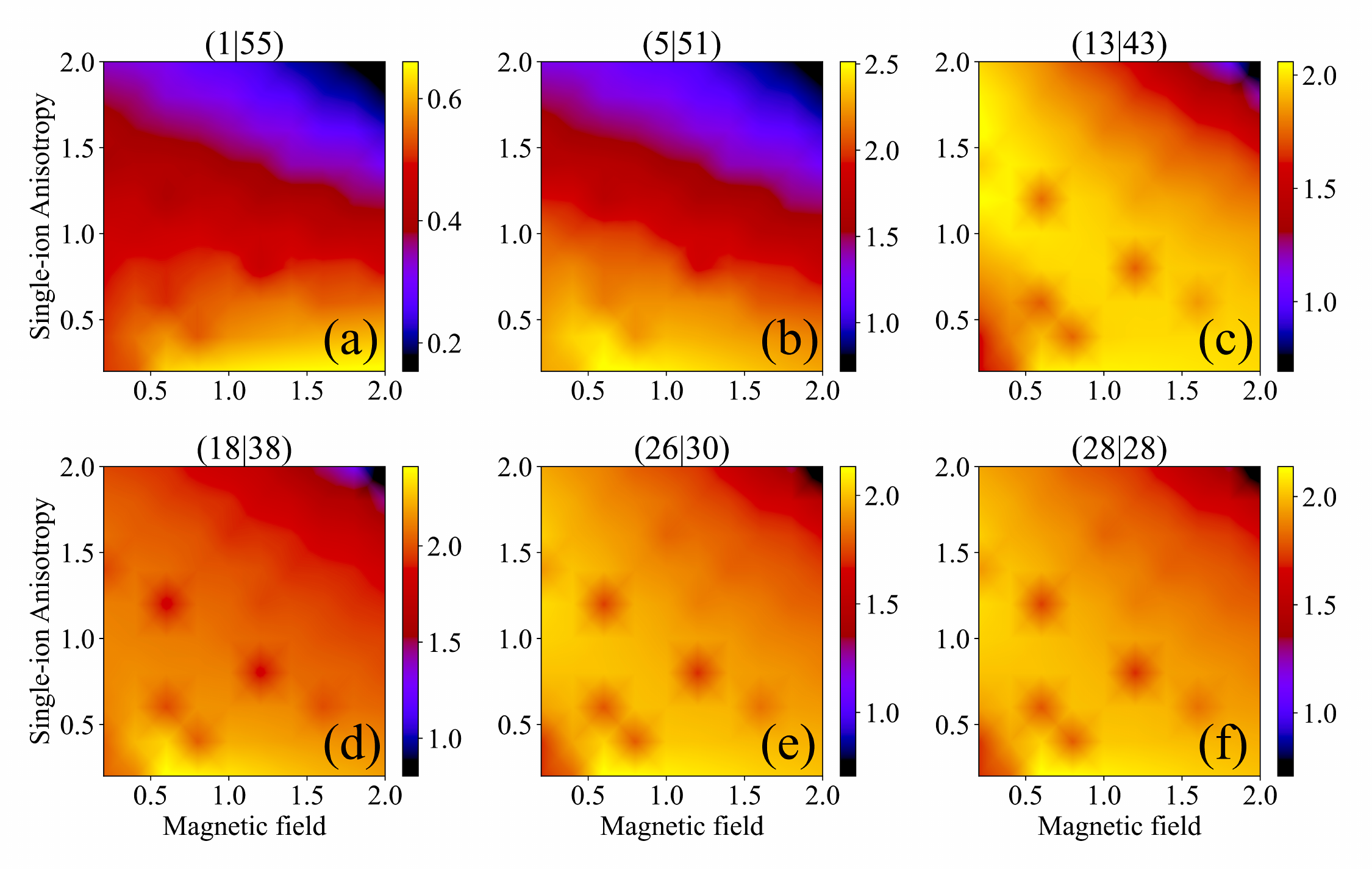}\\
\caption{The bipartite entanglement entropies were plotted as a function of magnetic field and anisotropy. The two numbers in the bracket $(M|N)$ refer to the two parties. The left party consists of the first $M$ spins, and the remaining spins are the right party. The entanglement entropy values are oscillating strongly for all the pairs as a function of $B$ and $D$.}\label{fig:ent2} 
\end{figure}

As shown in Fig.~\ref{fig:ent2}, the entanglement entropy was evaluated as a function of $B$ and $D$. In the notation $(M|N)$, $M$ and $N = 56 - M$ denote the bi-partition of the system: the first $M$ spins versus the remaining $N$. Across all partitions, for low anisotropy values ($D = 0$–$0.4$), the entanglement entropy initially increases with $B$, then decreases at higher fields. When the left party is on the edge, such as $(1|55)$ and $(5|51)$, the entropy decreases significantly with increasing $B$ and $D$, reaching a minimum when both are set to 2.

By contrast, more central partitions—$(13|43)$, $(26|30)$, and $(28|28)$ (Fig.\ref{fig:ent2}c, e, f)—exhibit non-monotonic behavior: the entropy first arises and then falls with increasing $B$ and $D$. This decline in entanglement reflects a change in the ground state—a quantum phase transition driven by the competition between the exchange interaction and the external parameters $B$ and $D$. Evidence of this transition is also seen in the expectation values of $\hat{s}_{i,x}$, $\hat{s}_{i,z}$, $\hat{T}_{i,x}$, and $\hat{T}_{i,z}$ shown in Fig.\ref{fig:sxsz}. For small $D$, the triplet spins are anti-aligned with surrounding $\frac{1}{2}$-spins , while for large $B$ and $D$, the edge spins become aligned.

This quantum phase transition is particularly clear in Fig.\ref{fig:ent2}(a–b), where strong $B$ and $D$ (top-right corners) indicate a shift from an anti-ferromagnetic (AFM) ground state at $B = D = 0$—between spin-$\frac{1}{2}$ and spin-1—to a ferromagnetic edge configuration at $B = D = 2$, while the bulk remains AFM, as shown in Fig.\ref{fig:sxsz}(a,c). This also explains why the peak in entanglement entropy shifts toward the bulk. Additionally, an oscillatory pattern in the entanglement entropy as a function of $B$ can be found in Fig.\ref{fig:ent2}, consistent with the previous results in Ref.~\cite{canovi2014dynamics}.

\begin{figure}[htbp]
\centering
\includegraphics[scale=0.24, trim={0.75cm 0.5cm 0.0cm 0.5cm},clip]{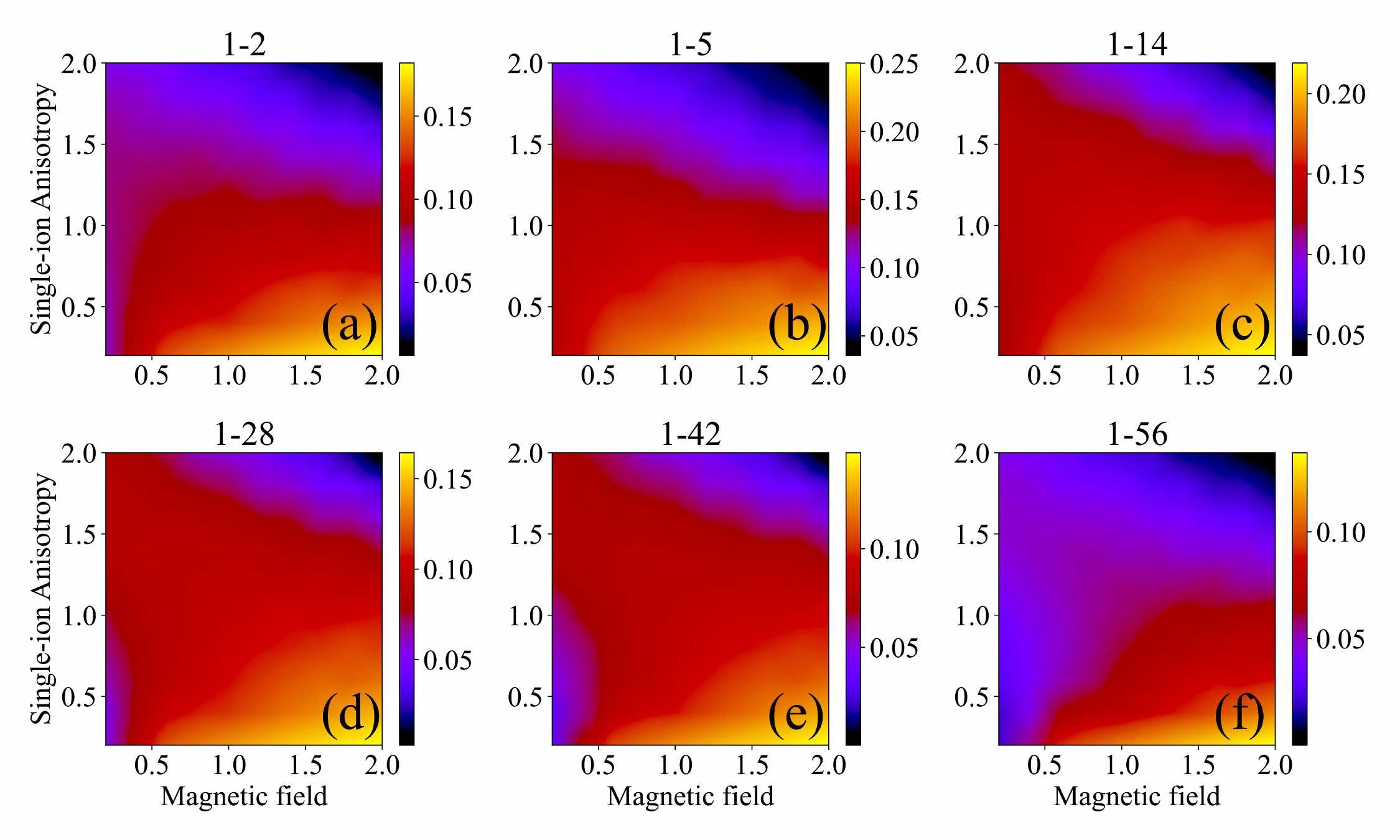}\\
\caption{The reduced density matrix elements for $\ket{\uparrow_a\downarrow_b}\bra{\downarrow_a\uparrow_b}$ map of $\rho_{a-b}$ are plotted as a function of magnetic field and anisotropy. The matrix elements are calculated for the different pairs of spins $a$ and $b$, i.e. (1,2), (1,5), (1,14), (1,28), (1,42) and (1,56). }\label{fig:rd2112} 
\end{figure}

Fig.\ref{fig:rd2112} shows the reduced density matrix elements $\bra{\uparrow_a\downarrow_b}\rho_{a-b}\ket{\downarrow_a\uparrow_b}$ for pairs of $\frac{1}{2}$-spins ($a$ and $b$), representing coherent spin-flip processes. These elements are evaluated for the spin pairs $(1,2)$, $(1,28)$, and $(1,56)$, corresponding to local, mid-range, and long-range coherence, respectively. All cases exhibit maximal coherence at low single-ion anisotropy ($D$) and high magnetic field ($B$), consistent with the entanglement entropy trends in Fig.~\ref{fig:ent2}.

The coherence for the $(1,2)$ pair reflects local spin dynamics, while the $(1,28)$ and $(1,56)$ pairs demonstrate significant long-range coherence. Remarkably, the coherence between the distant sites $(1,56)$ reaches values up to 0.15, indicating the persistence of long-range entanglement mediated by collective triplet excitations, even under moderate $B$ and $D$.

All matrix elements show enhanced values near $B = D = 1$, suggesting a strong competition between the exchange interaction and the external parameters. This is highlighted by the prominent red regions in Fig.\ref{fig:rd2112}. The observed coherence transitions align with the quantum phase transitions, as indicated by the entanglement entropy in Fig.\ref{fig:ent2}. In addition, the signs of the quantum phase transition behavior are visible along the $D$-axis when $B < 0.5$ (bottom-left region of Fig.~\ref{fig:rd2112}d–f), indicating the nuanced role of anisotropy even at low magnetic field strengths.

\begin{figure*}[htbp]
\centering
\includegraphics[scale=0.5, trim={0.5cm 0cm 0.0cm 0.0cm},clip]{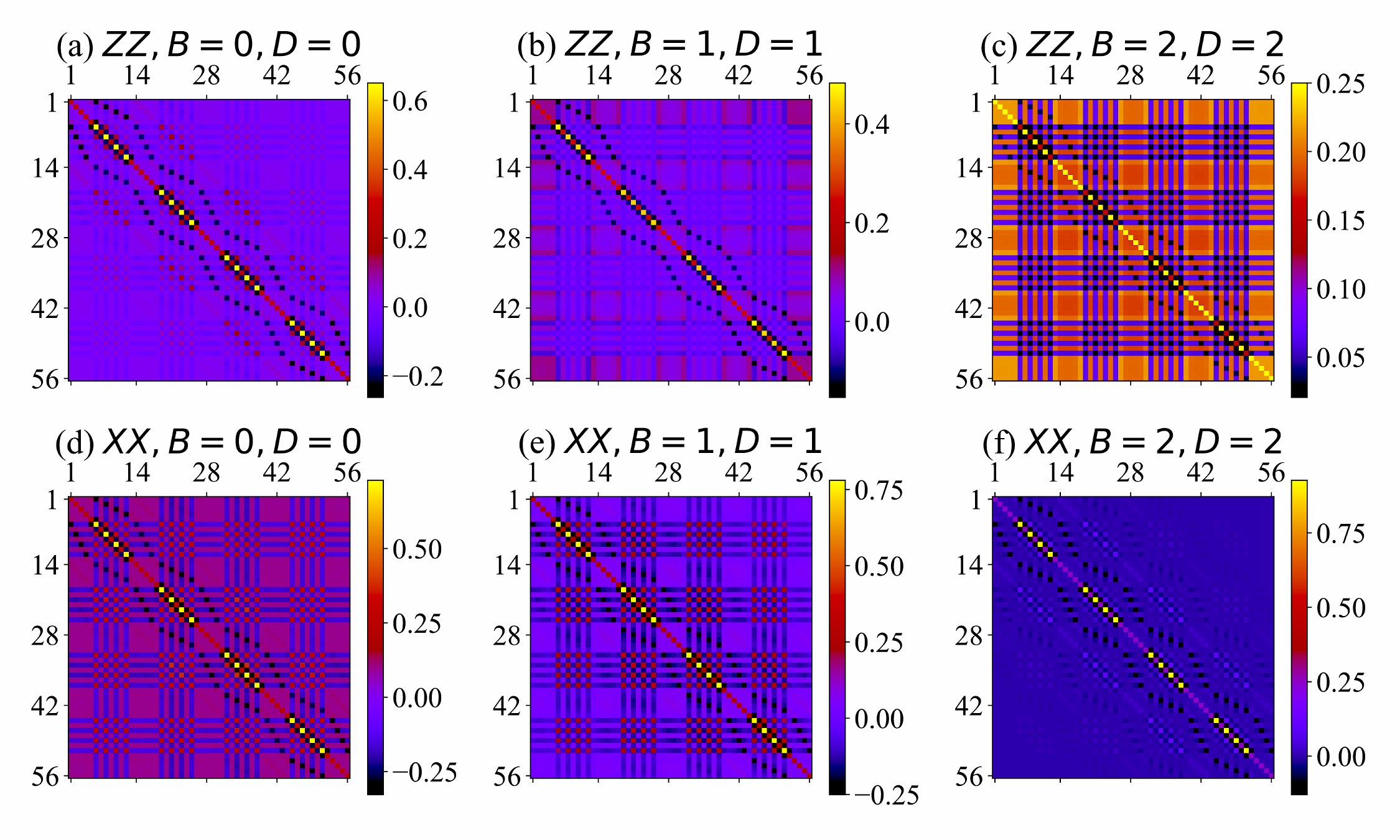}\\
\caption{The $XX$ and $ZZ$ spin-spin correlations for different magnetic fields and magnetic anisotropies are shown. These correlation functions are computed for $B=0,D=0$ (a,d), $B=1,D=1$ (b,e), and $B=2, D=2$ (c,f), respectively. }\label{fig:xxzz} 
\end{figure*}

Spin–spin correlation functions are shown in Fig.\ref{fig:xxzz}, for varying magnetic field strengths and single-ion anisotropies. The results indicate that the 16 triplets significantly influence the spin–spin correlations, as shown by the square patterns formed by them. As expected, strong magnetic fields and anisotropies (Fig.\ref{fig:xxzz}c,f) suppress these correlations. In contrast, both $XX$ and $ZZ$ spin-spin correlation functions demonstrate that the system remains coherent under weak magnetic field and anisotropy conditions (Fig.~\ref{fig:xxzz}a,b,d,e).

\section{Conclusions} 
The von Neumann entanglement entropy in a molecular quantum circuit driven by optically active triplet states—comprising 16 triplets and 40 $\frac12$-spins—has been computed and shows an edge mode for entanglement, mesoscopic entanglement and quantum phase transition. Spin coherence was characterized theoretically using reduced density matrix elements and spin–spin correlation functions. The dependence of entanglement and coherence on the external magnetic field and single-ion magnetic anisotropy, was systematically analyzed. When magnetic anisotropy is small, the magnetic field enhances the entanglement entropy, likely by introducing effective randomness. In contrast, single-ion magnetic anisotropy tends to suppress spin coherence. Both parameters tend to shift entanglement from the circuit edges toward the bulk. The computed reduced density matrix and the spin–spin correlation functions demonstrate a persistent long-range coherence across the system. Notably, varying the control parameters $B$ and $D$ reveals quantum phase transitions due to their competition with exchange interactions. These investigations offer new foundational insight into the entanglement structure of triplet-driven molecular quantum computing architectures. Future work will explore excited spin states of the quantum circuit, the time evolution of the Hamiltonian, and quantum gate tomography, particularly when selectively activating triplets within the circuit. 

\section{Hamiltonian and Methods} 
To entangle spins, we typically need to turn on the interactions between them. The interactions include exchange interaction (the strongest short-range interaction) and dipolar interaction that becomes strong at the long range. Magnetic anisotropy arising from the spin-orbit coupling is also an important parameter for the triplet.  \cite{huang2020emergent,zhang20242d,kurebayashi2022magnetism}.  
The Hamiltonian associated with the network in Fig.\ref{fig:molecularqc} therefore reads as follows.

\begin{equation}\label{eq:1}
    \hat{H} = \sum_{\vec{i},\vec{\delta}} [\hat{\vec{T}}_{\vec{i}}\cdot\hat{\vec{s}}_{\vec{i}+\vec{\delta}}]+\sum_{\vec{i},\vec{\delta}^\prime} \vec{B}\cdot(\hat{\vec{T}}_{\vec{i}}+\hat{\vec{s}}_{\vec{i}+\vec{\delta}^\prime})+D\hat{T}_{\vec{i},z}^2]
\end{equation}

$\hat{\vec{T}}$ and $\hat{\vec{s}}$ are spin-$1$ and spin-$\frac12$, respectively. $\vec{B}$ is the external magnetic field chosen to be along the $z$-direction. $D$ is the single-ion anisotropy for spin-$1$. Here $\vec{i}$ is used to label the site of the triplet,  $\vec{\delta}$ refers to the relative positions of the $\frac12$-spins to the coupled triplet and $\vec{\delta}^\prime$ removes the double counting and makes sure only counting spin-$\frac12$ once. So $\vec{\delta}$ and $\vec{\delta}^\prime$ can be equal to $(\pm1,0)$ and $(0,\pm1)$. The exchange interaction is set as $1$.

The tensor network (TN) method was employed to compute the ground state of the spin system \cite{orus2014practical, orus2019tensor, schollwock2011density,itensor,itensor-r0.3}. 
The ground state was calculated using the density matrix renormalization group (DMRG) method, expressed in the form of matrix product states (MPS), across various external magnetic field strengths and magnetic anisotropies applied to the triplet states \cite{schollwock2011density}. Spin operators are expressed using the matrix product operator (MPO) formalism.

The system was arranged into a one-dimensional configuration by ordering the sites as illustrated in Fig.~\ref{fig:molecularqc}. The MPS calculations employed a maximum bond dimension of 800 and ran 10 DMRG sweeps. The maximum truncation error from singular value decomposition (SVD) and density matrix diagonalizations was in the order of $10^{-7}$ after convergence. The von Neumann entanglement entropy was then computed using the SVD eigenvalues, as detailed by the following.
   $ S_{A|B}(\ket{\psi}) = -\text{Tr}\hat{\rho}_A\text{log}_2\hat{\rho}_A = - \sum_{a=1}^r s_a^2\text{log}_2s_a^2$.
Here $S_{A|B}$ is the von Neumann entanglement entropy, $\ket{\psi}$ is the ground state, $\hat{\rho}_A$ is the reduced density matrix for the \textit{A} party, and $s_a$ are the single values. The sum runs over all the non-vanishing single values. The reduced density matrix was computed as follows,
    $\rho_{s_i,s_j}^{s_i^\prime,s_j^\prime}=\text{Tr}_{s\neq s_i,s_j}[\ket{\psi}\bra{\psi}]$.
The expectation values of $\hat{s}_x$ , $\hat{s}_z$, $\hat{T}_x$  and $\hat{T}_z$ were computed for the ground state. The spin-spin correlations for different pairs of sites $(i,j)$, including $ C_{i,j}^{XX}= \bra{\psi}S_i^xS_j^x\ket{\psi}$ and $C_{i,j}^{ZZ}= \bra{\psi}S_i^zS_j^z\ket{\psi}
$, were also calculated for the ground state. 
This calculation can be performed in a straightforward manner within the MPS formalism. All the methods used here have been carefully benchmarked on the known smaller systems, including spin-$\frac12$ chains and a triplet-doublet pair. 

\section*{ACKNOWLEDGMENTS}
WW thanks Prof. Andrew Fisher (UCL), Prof. Nicholas Harrison (Imperial College London), Dr Kang Wang (Institute of Physics, Chinese Academy of Sciences, CAS IOP), Dr Haijun Liao (CAS IOP), and Prof. Tao Xiang (CAS IOP) for the inspiring and helpful discussions. 



\bibliography{spin2d}

\end{document}